\documentclass[11pt,english,aps]{revtex4}
\usepackage[T1]{fontenc}
\usepackage[latin1]{inputenc}
\usepackage{color}
\usepackage{graphicx}
\usepackage{amssymb}

\makeatletter

\usepackage{babel}
\makeatother
\begin{document}

\preprint{April 2006}

\title{Validity of the Rooted Staggered Determinant in the continuum limit}

\author{Anna Hasenfratz}

\email{anna@eotvos.colorado.edu}

\affiliation{Department of Physics, University of Colorado, Boulder, CO-80309-390}

\author{Roland Hoffmann}

\email{hoffmann@pizero.colorado.edu}

\affiliation{Department of Physics, University of Colorado, Boulder, CO-80309-390}

\begin{abstract}
We investigate the continuum limit of the rooted staggered determinant
in the 2-dimensional Schwinger model. We match both the unrooted and
rooted staggered determinant with an overlap fermion determinant of
two (one) flavors and a local pure gauge effective action by fitting
the coefficients of the effective action and the mass of the overlap
operator. The residue of this fit measures the difference of the staggered
and overlap fermion actions. We show that this residue scales at least
as $O(a^{2})$, implying that any difference, be it local or non-local,
between the staggered and overlap actions becomes irrelevant in the
continuum limit. This observation justifies the rooting procedure.
\end{abstract}
\maketitle

\section{Introduction}

Staggered fermions offer many computational advantages over other
fermion formulations. Simulations can be performed in large volumes
at fairly small quark masses and data with improved actions show small
scaling violations. However, the staggered action does not have full
chiral symmetry and the chiral limit has to be taken together with
the continuum limit. This is no different from other non-chiral actions,
but staggered fermions have another, potentially serious problem.
In 4 dimensions the staggered action describes four species (or tastes)
of fermions, it cannot describe a single Dirac particle directly.
In order to reduce the number of tastes from four to one the 4th root
of the fermion determinant is taken in the path integral and there
is no a priori reason that this rooted determinant corresponds to
a local fermionic action belonging to the same universality class
as 1--flavor QCD.

Several analytical and numerical works addressed this question in
the last few years \cite{Bunk:2004br,Adams:2004mf,Maresca:2004me,Shamir:2004zc,Durr:2003xs,Neuberger:2004be,Durr:2004ta,Bernard:2004ab,Bernard:2005gf,Hasenfratz:2005ri}.
None of them showed evidence that the procedure introduces non--universal
errors, i.e. errors that cannot be considered cutoff effects that
scale away in the continuum limit, but neither could they prove the
validity of the rooting procedure. Recently it has been argued, based
on a number of reasonable conjectures, that while the rooted staggered
action is non-local at any finite lattice spacing, in the continuum
limit the non-local terms become irrelevant \cite{Bernard:2006xx,Shamir:2006xx}.

In this paper we present numerical evidence obtained in the 2--dimensional
Schwinger model, showing that the rooted staggered action is in the
right universality class. While the Schwinger model is much simpler
than 4--dimensional QCD, its basic properties are QCD--like and therefore
we believe that this work gives strong indication that the rooting
procedure is safe and valid for 4--dimensional QCD simulations. \textcolor{blue}{}\textcolor{black}{We
also show that the staggered action can be considered equivalent to
a chiral Ginsparg-Wilson action only when the staggered mass is larger
than typical taste symmetry breaking effects, limiting the parameter
space where staggered simulations can be expected to approximate continuum
QCD. We describe how the masses of the staggered and corresponding
overlap actions should be matched to obtain physically equivalent
theories when this condition is satisfied.}

\section{The continuum limit of the staggered action\label{sec:The-continuum-limit}}

The partition function of the unrooted staggered action is \begin{eqnarray}
Z & = & \int\!\!\! D[U\bar{\psi}\psi]\, e^{-S_{_{g}}(U)-\bar{\psi}(M+am_{{\rm {st}}})\psi}\label{Part.Func.}\\
 & = & \int\!\!\! D[U]\,{\rm {det}}(M+am_{{\rm {st}}})\, e^{-S_{g}(U)}\,,\nonumber \end{eqnarray}
 where $S_{g}(U)$ is a gauge action, $M$ is the staggered Dirac
operator and $am_{{\rm {st}}}$ is the bare staggered mass. The staggered
action has two relevant couplings, a gauge coupling that determines
the lattice spacing $a$ and the fermion mass $m_{{\rm {st}}}$. In
the $a\to0$ continuum limit the staggered action describes $n_{t}=4$
degenerate fermions in 4, $n_{t}=2$ fermions in 2 dimensions. At
finite lattice spacing the taste symmetry is broken, the action describes
$n_{t}$ fermion tastes but only with a remnant $U(1)$ taste symmetry.%
\begin{figure}
\includegraphics[%
  scale=0.7]{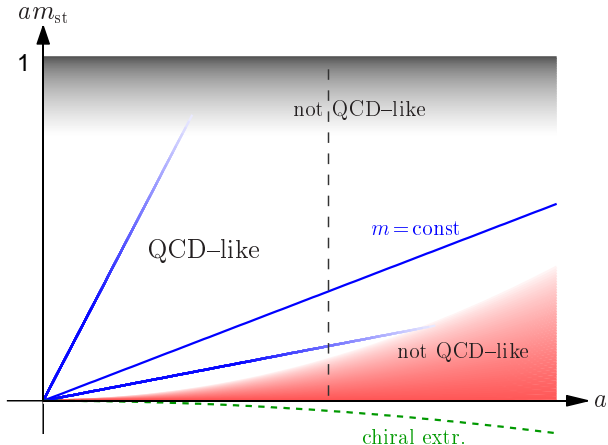}

\caption{\label{fig:phase-diag}The expected phase diagram of the unrooted
staggered action. The solid blue lines show how the continuum limit
is approached with fixed physical mass. This approach should avoid
the shaded regions dominated by cutoff effects or strong taste symmetry
violation, respectively. Neither of those is expected to show QCD--like
behavior. }
\end{figure}
The phase diagram in the relevant parameters $a$ and $am_{{\rm st}}$
is sketched in Fig.\ref{fig:phase-diag}, where the solid blue lines
illustrate how the continuum limit with fixed  mass can be approached. 

Fixed finite lattice spacing corresponds to a vertical line, like
the dashed line in Fig.\ref{fig:phase-diag}. The latter can be divided
into three regions:

\begin{itemize}
\item At $am_{st}=0$ the staggered action's spectrum has a single Goldstone
particle and $n_{t}^{2}-1$ massive pseudoscalars. While $n_{t}^{2}-2$
of these will become massless as $a\to0$, at any finite lattice spacing
the staggered spectrum is very different from $n_{t}$--flavor massless
QCD. At small fermion mass $am_{{\rm {st}}}\gtrsim0$ the taste breaking
terms dominate and the non--Goldstone pions are heavy compared to
the Goldstone one. Again one does not expect QCD--like behavior.
\item $am_{{\rm {st}}}\gtrsim1$ is the cutoff region (upper shaded area),
again not continuum QCD--like.
\item Only in the middle of the diagram, between the two shaded regions,
would one expect to observe QCD. The $a\to0$, $am_{{\rm {st}}}\to0$
continuum limit should be approached here.
\end{itemize}
While staggered fermions formally allow $am_{{\rm {st}}}=0$, physically
this limit does not correspond to QCD at any finite lattice spacing
\cite{Durr:2004ta,Bernard:2004ab}. Simulations cannot be trusted
at a small fermion mass where taste breaking terms dominate the pseudoscalar
sector. However, the taste breaking terms are expected to scale at
least with $O(a^{2})$, such that at small enough lattice spacing
the continuum limit can be approached with any finite fermion mass.
Thus the exclusion of $am_{{\rm {st}}}=0$ is not a serious problem
for massive fermions. We will discuss the case of massless fermions
further at the end of Sect. \ref{sec:Schwinger-model}.

The staggered determinant can always be written as \begin{equation}
{\rm det}(M+am_{{\rm {st}}})=\mathrm{det}{}^{n_{t}}(D_{{\rm {1f}}}+am_{{\rm {1f}}}){\rm \,{det}}(T)\,,\label{Det_relation}\end{equation}
 where $D_{{\rm {1f}}}+am_{{\rm {1f}}}$ is an arbitrary 1--flavor
Dirac operator and ${\rm det}(T)$ describes all the terms that are
not included in the latter. If the local $D_{{\rm {1f}}}$ operator
and the mass term $m_{{\rm {1f}}}$ could be chosen such that $T$
contains only local gauge terms,\begin{equation}
{\rm {det}}(T)=e^{-S_{{\rm {eff}}}(U)}\,\,,\label{local_T}\end{equation}
 the staggered action would differ from an $n_{t}$--flavor degenerate
Dirac operator only in cutoff level terms \cite{Adams:2004mf}. This
is indeed the case for heavy, $am_{{\rm {st}}}\!\gtrsim\!1$ fermions,
the upper shaded region in Fig.\ref{fig:phase-diag}. 

On the other hand there are several examples that illustrate that
at $am_{{\rm {st}}}=0$ the operator $T$ cannot be local at any finite
lattice spacing. In 4 dimensions the staggered theory has one massless
Goldstone boson and 15 heavy pseudoscalars at vanishing bare quark
mass. A theory with 4 degenerate flavors has either 15 massless Goldstone
bosons if $am_{{\rm {1f}}}=0$ or none if $am_{{\rm {1f}}}\ne0$.
None of these possibilities can match the spectrum of the staggered
theory \cite{Bernard:2006xx}. Similarly, on topologically non--trivial
configurations a chiral $D_{{\rm {1f}}}$ has an exact zero mode per
topological charge, ${\rm {det^{4}}(}D_{{\rm {1f}}}+am_{{\rm {1f}}})$
vanishes as $am_{{\rm {1f}}}\to0$. The staggered operator does not
have exact zero modes at finite lattice spacing and therefore the
left hand side of Eq.(\ref{Det_relation}) is finite even when $am_{{\rm {st}}}=0$.
If $am_{{\rm {st}}}=0$ implies $am_{{\rm {1f}}}=0$, ${\rm {det}}(T)$
has to diverge, it cannot be identical to a local gauge operator at
finite lattice spacing \cite{Hasenfratz:2005ri,Bernard:2006xx} %
\footnote{The inconsistency of (\ref{Det_relation}) on topologically non--trivial
configurations was observed in Ref. \cite{Hasenfratz:2005ri}. In
an attempt to keep the operator $T$ local, Ref. \cite{Hasenfratz:2005ri}
suggested a mass shift between the staggered and overlap actions.
Based on the results presented here and in Ref. \cite{Bernard:2006xx}
we now conclude that $T$ is non--local in the physically relevant
regime even when allowing for a mass shift at finite lattice spacing.%
}. Numerical simulations of the finite temperature phase transition
at zero fermion mass in 4 dimensions also indicate that the massless
theory is not QCD--like \cite{Kogut:2006gt}, showing $O(2)$ rather
then $O(4)$ critical exponents.

Recently Bernard at al. \cite{Bernard:2006xx} argued that $\mathrm{det}(T)$
cannot be a local operator even at finite fermion mass at finite lattice
spacing. This, however, does not mean that the staggered operator
cannot describe QCD in the continuum limit. If we write the determinant
as \begin{equation}
{\rm {det}}(T)=e^{-S_{{\rm eff}}(U)}\,{\rm {det}}(1\!+\!\Delta)\,,\label{non-local_T}\end{equation}
and can choose $S_{{\rm {eff}}}$ such that the non--local term $\Delta$
is bounded at finite mass and goes to zero as $a\to0$, the staggered
determinant in Eq.(\ref{Det_relation}) will describe $n_{t}$ degenerate
flavors in the continuum limit. This is certainly the expected behavior
for the unrooted action.

Now we turn our attention to the rooting procedure. With the notation
introduced above the root of the staggered determinant is \begin{equation}
\mathrm{det}^{1/n_{t}}(M+am_{{\rm {st}}})={\rm {det}}(D_{{\rm {1f}}}+am_{{\rm {1f}}})\, e^{-S_{{\rm eff}}(U)/n_{t}}\,{\mathrm{det}}^{1/n_{t}}(1\!+\!\Delta).\label{rooted_det}\end{equation}
If one could show that a\begin{eqnarray}
\Delta\to0\,\, & {\rm {as}} & a\to0\,,\label{required}\end{eqnarray}
 the rooted determinant of Eq.(\ref{rooted_det}) would correspond
to a local 1--flavor action in the continuum limit. 

Based on renormalization group arguments, in Ref. \cite{Shamir:2004zc}
Shamir showed that this is indeed the case for free fermions. In a
recent work \cite{Shamir:2006xx}, based on a number of reasonable
assumptions, he argues that the same is true in the interacting theory. 

Refs. \cite{Shamir:2004zc,Shamir:2006xx} describe an explicit method
to find $D_{{\rm {1f}}}+am_{{\rm {1f}}}$ and the corresponding effective
gauge action $S_{{\rm {eff}}}$, but the construction is not unique.
In fact, any Dirac operator and effective gauge action that satisfy
Eqs.(\ref{rooted_det},\ref{required}) will justify the rooting procedure.
In the following we pick an arbitrary Ginsparg-Wilson  operator as
$D_{{\rm {1f}}}$ and ask if $am_{{\rm {1f}}}$ and $S_{{\rm {eff}}}(U)$
can be chosen such that Eq.(\ref{required}) is satisfied.

\section{Matching the fermionic determinants}

\textcolor{black}{The actual matching strategy is fairly general and
we will describe it for an arbitrary pair of Dirac operators $D_{1}+am_{1}$
and $D_{2}+am_{2}$. We want to know to what extent the determinant
of the first Dirac operator can be described by the determinant of
the second plus pure gauge terms. To find this we calculate the determinant
ratio\begin{equation}
\mathrm{det}(T)=\frac{\mathrm{det}(D_{1}+am_{1})}{\mathrm{det}(D_{2}+am_{2})}\label{det_ratio}\end{equation}
 on a set of dynamical configurations generated at some given lattice
spacing with the action \begin{equation}
S_{1}=S_{g}(U)+\bar{\psi(}D_{1}+am_{1})\psi\,.\label{S_dyn}\end{equation}
 Next we fit the logarithm of the determinant ratios with a pure gauge
action of the form \begin{equation}
S_{{\rm {eff}}}=\sum_{l=0}^{l=n}\alpha_{l}\,\mathcal{C}_{l}(U)\,,\label{S_eff}\end{equation}
where $\mathcal{C}_{l}(U)$ denotes traces of Wilson loops.} The effective
action $S_{{\rm {eff}}}$ has to be local, the coefficients $\alpha_{l}$
have to decay exponentially with the length of the loops. In practice
we use an ultralocal effective action. \textcolor{black}{The accuracy
of the matching at fixed fermion mass $m_{2}$ is characterized by
the per flavor/taste residue }%
\footnote{In the staggered case we use $(M+am_{{\rm {st}}})^{1/n_{t}}$ at this
point, which reduces the coefficients $\alpha_{l}$ and the residue
$r$ by a factor $n_{t}$, thus giving the per flavor residue.%
}\textcolor{black}{\begin{equation}
r(m_{2})=\Big\langle\Big({\rm {log}\,}\frac{{\rm {det}}(D_{1}+am_{1})}{{\rm {det}}(D_{2}+am_{2})}-S_{{\rm {eff}}}(U)\Big)^{2}\Big\rangle^{1/2}\,.\label{residue}\end{equation}
The minimum of the residue $r(m_{2})$ in terms of $m_{2}$ determines
the action $D_{2}+am_{2}$ that is} \textcolor{black}{\emph{physically
closest}} \textcolor{black}{to the original} $D_{1}+am_{1}$ action.
In this sense it defines the mass $\overline{m}_{2}$ that matches
the fermion mass $m_{1}$. In the notation of Eq.(\ref{non-local_T})
then \begin{equation}
r(\overline{m}_{2})=\Big\langle\Big(({\rm {log\, det}}(1\!+\!\Delta)\Big)^{2}\Big\rangle^{1/2}\,.\label{matched_r}\end{equation}
If the two fermion operators describe the same continuum theory the
residue has to vanish as $a\to0$ at fixed volume and quark mass.

\section{Schwinger model - numerical results \label{sec:Schwinger-model}}

\subsection{Setup and matching tests }

The 2--dimensional Schwinger model offers an excellent testing ground
for the matching idea as it can be studied with high accuracy and
limited computer resources. It is a super--renormalizable theory since
the bare gauge coupling $g$ is dimensional, the lattice gauge coupling
is $\beta=1/(ag)^{2}$. A continuum limit in fixed physical volume
can be achieved by keeping the scaling variable $z=Lg$ fixed while
increasing the lattice resolution. We choose $z=6$ and vary the lattice
size between $L/a=12$ and $L/a=28$ so the corresponding gauge coupling
$\beta=(L/a)^{2}/z^{2}$ varies between $4$ and $\sim22$, in or
at least close to the the scaling regime. The scaling parameter $z$
characterizes the (physical) volume while we use $mL$ to fix the
mass. 

We produced gauge configurations using a global heatbath for the plaquette
gauge action. The use of a global algorithm is essential to this study
since at large values of $\beta$ the autocorrelation time for topological
charge fluctuations increases dramatically with both (local) heatbath
and Metropolis. In fact, at the chosen physical volume these algorithms
no longer tunnel between topological sectors at all for $\beta\gtrsim10$
and thus in practice lose ergodicity.

In the data analysis, measurements on the pure gauge ensemble are
reweighted with the appropriate power of the fermion determinant to
obtain the observables in the full dynamical theory. On the gauge
configurations we measure a set of Wilson loops $\mathcal{C}_{l}$
as well as the complete spectra of the Dirac operators under consideration.
For the matching we use an effective action (see Eq.\ref{S_eff})
that contains 9 loops up to length 10. With a maximal extension of
four lattice units $S_{{\rm eff}}$ is very localized even on our
coarsest lattices and in particular we do \emph{not} increase the
size or number of loops as we approach the continuum. Naturally, these
Wilson loops are strongly correlated and in all cases a similar quality
matching could be achieved using 3 appropriately chosen loops only.
On the other hand, the matching does not improve significantly when
using many more loops in the fit, indicating that Eq.(\ref{local_T})
cannot be satisfied.

As a first test of our matching method, we compared a smeared overlap
action to an unsmeared overlap action. Both actions have the same
(plain Wilson) kernel but different Ginsparg-Wilson $R_{0}$ parameters.
As expected in the case where both operators respect chiral symmetry,
we see a linear relation between $m_{1}$ and the matching $\overline{m}_{2}$
for small quark masses and the residue $r(\overline{m}_{2})$ is very
small ($3\%$ at $\beta=4$, decreasing to $0.2\%$ on the finest
lattice). Moreover, there is no observable dependence of $r(\overline{m}_{2})$
on the fermion mass $m_{1}$, which implies that the matching we attempt
works equally well at all masses. Next we matched a Wilson Dirac operator
with a smeared overlap action. We found that the residue is significantly
larger ($20\%$ decreasing to $4\%$) but scales as expected when
$a$ is decreased. Also, by extrapolating the matched Wilson mass
to vanishing overlap mass, we predict the critical mass $am_{{\rm c}}$
in excellent agreement with the large volume results of Ref. \cite{Christian:2005yp}
that used the pion mass to identify $am_{c}$. This agreement shows
that our matching procedure, while not based on physical observables,
identifies the parameters where the two actions are physically closest.

In the following we concentrate on the matching of the staggered action.
As a matching action we use a smeared overlap action with Ginsparg-Wilson
radius $R_{0}=1$ and plain $r=1$ Wilson kernel. The smearing is
a single (projected) APE step with $\alpha=0.4$ smearing parameter.
The specific choice of the matching overlap action is not particularly
important but we found better matching with the smeared link action.
We have done exploratory studies with unsmeared overlap actions and
with different $R_{0}$ choice but the results were not significantly
different.

\subsection{The unrooted staggered action}

We start our investigation with the unrooted action, which in 2 dimensions
corresponds to two fermion tastes. In the continuum limit it is expected
to describe two degenerate flavors and thus it should differ from
a degenerate 2--flavor overlap action in irrelevant terms only. Our
goal is to match the staggered determinant with a 2--flavor overlap
action and local gauge terms at fixed physical mass and volume and
study the residue of the matching (Eqs.\ref{residue},\ref{matched_r})
as the lattice spacing $a\to0$. Since we do not expect any strange
behavior here, this section serves as an illustration and test of
the matching procedure. 

\begin{figure}
\includegraphics[%
  scale=0.7]{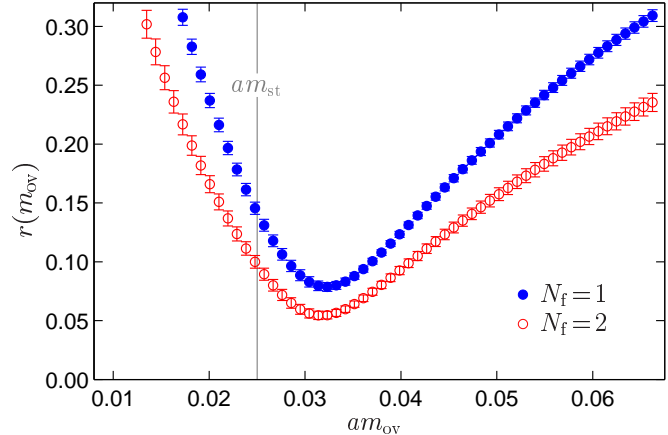}

\caption{The residue of Eq.(\ref{residue}) as the function of the matching
mass. The dynamical $L/a=20$ configurations were generated with the
staggered action at $am_{{\rm {st}}}=0.025$ and matched with an overlap
action. Open red circles: 2 tastes/flavors; filled blue dots: 1 taste/flavor.
\label{fig:minimum}}
\end{figure}
 Fig.\ref{fig:minimum} shows the matching of the $n_{t}=2$ staggered
determinant with the $N_{f}=2$ flavor-degenerate overlap determinant
at $z=6$ on $L/a=20$ lattices ($\beta\simeq11.11$). The quenched
configurations were reweighted to the dynamical staggered ensemble
at $am_{{\rm {st}}}=0.025$. The residue of the matching (Eq.\ref{residue})
has a well defined minimum at $a\overline{m}_{{\rm ov}}=0.0317(3)$. 

\begin{figure}
\includegraphics[%
  scale=0.7]{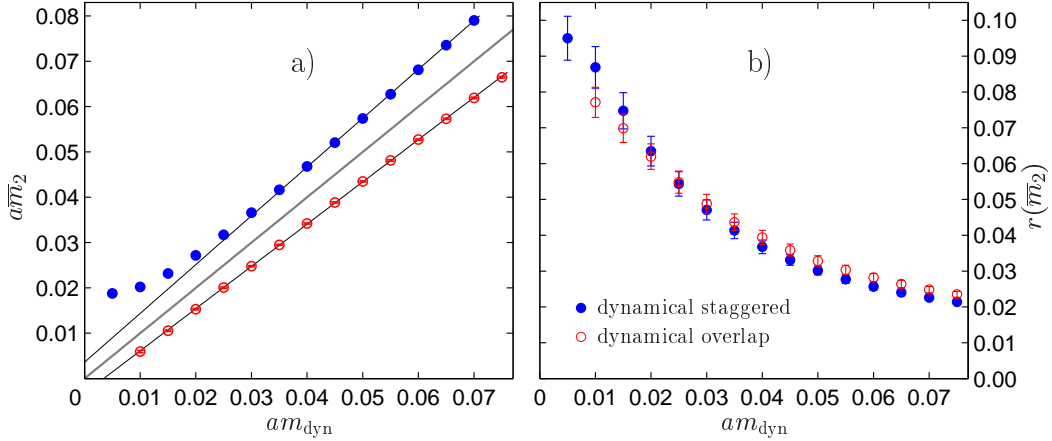}

\caption{a) The matching mass as the function of the dynamical action mass
at fixed lattice spacing for 2 tastes/flavors. Filled blue dots: staggered
action matched with overlap; open red circles: overlap action matched
with staggered. b) The residue of the matching described in a). \label{fig:Chiral}}
\end{figure}

By repeating the matching at different values of the staggered masses
$am_{{\rm {st}}}$ we can find the matching overlap masses at the
given lattice spacing as shown by the blue dots in Fig.\ref{fig:Chiral}a.
For larger masses the data show a linear dependence with a constant
offset, $\overline{m}{}_{{\rm {ov}}}=1.077(7)m_{{\rm {st}}}+0.0036(4)/a$.
This kind of functional form was conjectured in Ref. \cite{Hasenfratz:2005ri}.
For small masses, below $am_{\mathrm{st}}\approx0.02$, there is a
clear deviation from the linear behavior. The residue of the fit also
shows a rapid increase below this value (Fig.\ref{fig:Chiral}b) indicating
that the matching is no longer meaningful. According to the discussion
in Sect.\ref{sec:The-continuum-limit} we interpret this as the staggered
action being QCD--like for $am_{\mathrm{st}}{\rm \gtrsim}0.02$ and
not QCD--like below. 

\begin{figure}
\includegraphics[%
  scale=0.7]{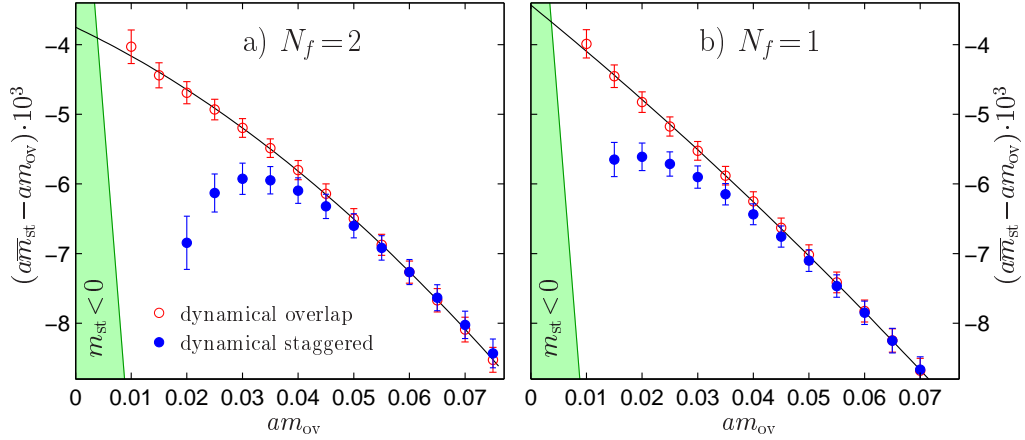}

\caption{The difference of the matched staggered and overlap masses vs. the
overlap mass at fixed lattice spacing. The notation is the same as
in Fig.\ref{fig:Chiral}. a) unrooted staggered/2-flavor overlap;
b) rooted staggered/1-flavor overlap matching.\label{fig:Regular}}
\end{figure}

As a consistency check we repeated the matching using the overlap
action for the dynamical configurations and matching the overlap determinant
with the staggered one. The result, shown by the red circles in Fig.\ref{fig:Chiral}a,
is the mirror image of the staggered with overlap matching data up
to the point where the latter matching breaks down. This is the expected
behavior if the two actions differ only by lattice artifacts. The
agreement is even more obvious in Fig.\ref{fig:Regular}a where we
replot the data of Fig.\ref{fig:Chiral}a showing the difference of
the matched staggered and overlap masses as the function of the overlap
mass. The only difference between the two data sets in Fig.\ref{fig:Chiral}
is the configuration ensemble used: staggered dynamical configurations
for the staggered with overlap matching, dynamical overlap configurations
for the overlap with staggered matching.

By restricting the configurations to the sector of trivial topology
we could verify that the difference between the matching on the two
ensembles and also most of the residue can be ascribed to configurations
with non--vanishing topological charge. The massless overlap operator
has a zero mode on these configurations while the smallest eigenvalue
of the staggered operator is non-zero, determined by the taste breaking
of the staggered action. Thus these configurations are not sufficiently
suppressed at small quark masses in the staggered ensemble. Our matching
procedure tries to compensate for this by assigning an even smaller
quark mass until the matching breaks down entirely. This effect should
become smaller when the number of flavors is reduced since the suppression
of topology becomes weaker also for the overlap ensemble.

The shaded area in Fig.\ref{fig:Regular} corresponds to the inaccessible
region of $am_{{\rm {st}}}<0$. The overlap with staggered matching
works basically up to this region. Extrapolating to the chiral limit
of the overlap action suggests that the critical mass for staggered
fermions is negative, $(am_{{\rm {st}}})^{{\rm cr}}=-0.0037(2)$ at
these parameter values. Of course this is a non-physical value, well
beyond the QCD like regime of staggered fermions. As argued above,
the staggered with overlap matching breaks down at a much larger mass.
Apparently configurations created with the staggered action at smaller
mass values cannot be reasonably described by an overlap action, signaling
the non QCD--like region of staggered fermions. %

\begin{figure}
\includegraphics[%
  scale=0.7]{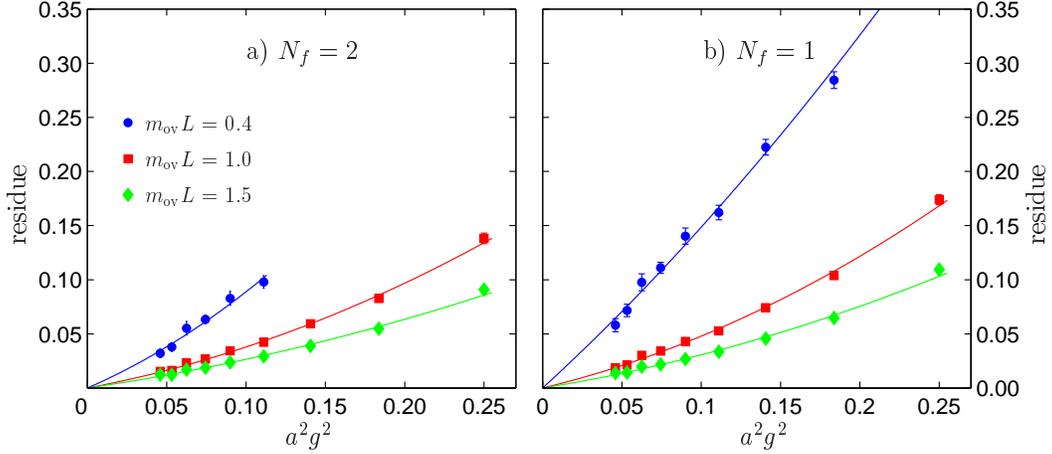}

\caption{Residue of the matching as a function of the (squared) lattice spacing
at different physical masses. a) unrooted staggered/2-flavor overlap;
b) rooted staggered/1-flavor overlap matching.\label{fig:Residue}}
\end{figure}

Next we consider the continuum limit of the matching at fixed physical
mass %
\footnote{We fix the physical mass by keeping $m_{{\rm ov}}L$ constant and
vary the staggered sea quark mass to achieve the matching.%
} and volume $z=gL=6$. Fig.\ref{fig:Residue}a shows the residue of
matching the 2--taste staggered determinant with the 2--flavor overlap
determinant at different masses as a function of $a^{2}g^{2}$. The
smaller the mass, the larger the matching residue is, as it is also
evident from Fig.\ref{fig:Chiral}b. For the smallest mass, $m_{{\rm ov}}L=0.4$
the data stops around $a^{2}g^{2}=0.11$ - on coarser lattices the
two actions cannot be matched, the residue of Eq.(\ref{residue})
has no minimum. Nevertheless matching is possible at smaller lattice
spacing and the residue at fixed $m_{{\rm ov}}L$ approaches zero
at least quadratically in $a$. The continuum limit can be approached
with any fermion mass and the staggered determinant can be described
as a 2--flavor chiral determinant plus pure gauge terms. This is the
behavior we expected from universality. %

\subsection{The rooted staggered action}

Now we repeat the analysis of the previous section for the rooted
staggered action. The procedure is very similar. The pure gauge configurations
are reweighted with the square root of the staggered determinant to
generate configurations with one taste and the rooted determinant
is matched with the 1-flavor overlap determinant plus pure gauge terms,
according to Eq.(\ref{residue}).

The quality of the matching is very similar to the unrooted/2--flavor
case as the open circles in Fig.\ref{fig:minimum} show. In fact,
even the matched mass $a\overline{m}_{{\rm ov}}=0.0322(2)$ hardly
differs from the 2--flavor case. The 1--flavor data in Figs.\ref{fig:Chiral}
is indistinguishable from the shown 2--flavor data. Fig.\ref{fig:Regular}b
shows the mass difference for the rooted/1--flavor matching. The matching
of the overlap determinant with staggered is almost the same as in
the unrooted case and also the ''critical'' mass $(am_{{\rm st}})^{{\rm cr}}=-0.0034(2)$
is in agreement with the 2--flavor data. As expected, the opposite,
i.e. the matching of the staggered determinant on staggered dynamical
configurations, works in a larger mass range in the rooted case.

Fig.\ref{fig:Residue}b is the important plot for the rooted staggered
action as it shows the residue at fixed physical masses as the continuum
limit is approached. There is a remarkable similarity between the
2-- and 1--flavor cases. \textcolor{blue}{}The residue for the 1--flavor
rooted determinant is larger but the continuum approach is identical,
at least quadratic in $a$. The taste violating term $\Delta$ in
Eqs.(\ref{non-local_T}) and (\ref{matched_r}) becomes irrelevant
in the continuum limit. This result justifies the rooting procedure.

\subsection{The phase diagram revisited}

\begin{figure}
\includegraphics[%
  scale=0.7]{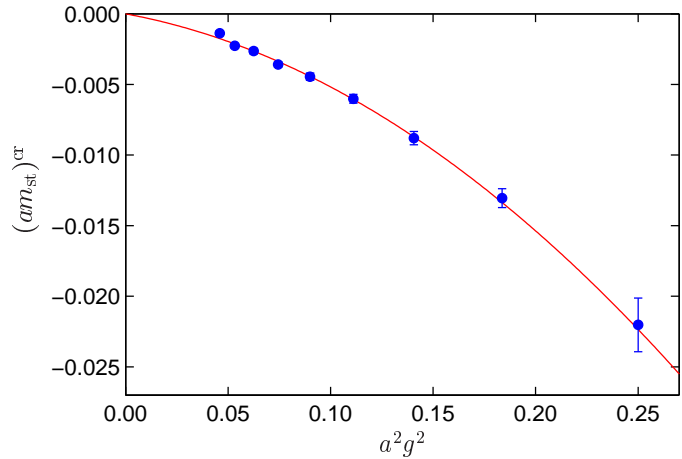}

\caption{The staggered mass that corresponds to a chiral overlap action as
the function of $a^{2}$. \label{fig:mcrit}}
\end{figure}

Our matching data can be used to quantify the phase diagram we sketched
in Fig.\ref{fig:phase-diag}. We have already discussed the chiral
extrapolation line, the staggered mass line that corresponds to the
chiral limit of the overlap action. While this {}``critical mass''
is not in the physically accessible region, it influences the relation
between the matched staggered and overlap masses. Its value is important
in mixed action simulations where an overlap valence quark action
is used with configurations generated with staggered sea quarks. Obviously
the deviation of this {}``critical mass'' from zero is a lattice
artifact and should vanish in the continuum limit. That is indeed
the case as is shown in Fig.\ref{fig:mcrit}, where we show $(am_{{\rm st}})^{{\rm cr}}$
from the 1--flavor data, which agrees with the unrooted one but has
smaller errors since the matching works closer to the chiral limit
(see above). %

\begin{figure}
\includegraphics[%
  scale=0.7]{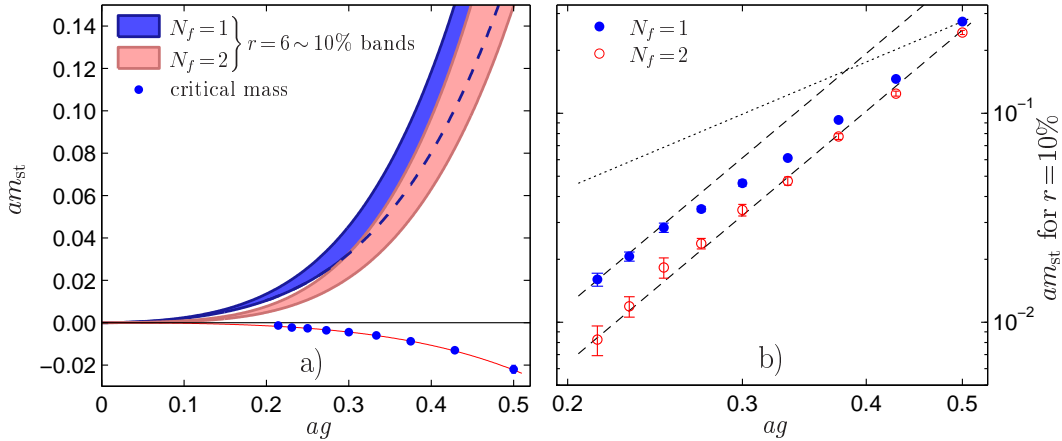}

\caption{a) The phase diagram of the staggered action. The bands denote the
regions where the matching residue is between $6$ and $10\%$, indicating
the onset of non QCD--like behavior. The data points at negative mass
show where the staggered action would correspond to a chiral overlap
action. b) Logarithmic plot of the staggered mass at $10\%$ matching
residue from part a).\label{fig:scaling}}
\end{figure}

We can also map out the QCD--like and non QCD--like regions as indicated
in Fig.\ref{fig:phase-diag}. While this is not a uniquely defined
boundary, its meaning is yet quite clear. To quantify it we adopt
a somewhat arbitrary but reasonable definition: we consider the matching
between the staggered and overlap actions possible if the matching
residue $r(\overline{m}_{{\rm ov}})$ (per flavor) is smaller than
some predetermined number. The shaded bands in Fig.\ref{fig:scaling}a
correspond to residues between $6\%$ and $10\%$. The darker blue
region is for the rooted/1--flavor case, the overlapping lighter red
band is the unrooted 2--flavor boundary. As expected, the 2--flavor
band approaches zero at least quadratically (see below). The interesting
thing is that the rooted 1-flavor data show an almost identical behavior,
again signaling that in the continuum limit the rooted action is as
good as the unrooted one.%

\subsection{Final comments\label{sub:Final-comments}}

The data we presented in the previous sections correspond to thin
link staggered fermions. Smeared link fermions have smaller taste
breaking and show better scaling. When repeating the analysis with
smeared staggered fermions, we found very similar behavior both for
the rooted and for the unrooted case but indeed with greatly reduced
taste breaking effects. In particular, at a given lattice spacing
the matching worked down to much smaller quark masses. The lowest
staggered eigenmode(s) on topologically non--trivial configurations
are significantly smaller after smearing and thus the dynamical suppression
of topology is much closer to that of a chiral overlap action.

Our data show that there is no problem with approaching the continuum
limit at fixed physical mass (Fig.\ref{fig:Residue}). Whether massless
fermions can be described in the continuum limit with the staggered
action depends on the actual scaling of the taste violating terms.
To achieve a chiral continuum limit the fixed point has to be approached
at least like $am_{{\rm st}}\propto a^{2}$, such that the physical
mass $\sim\partial(am_{{\rm st}})/\partial a$ vanishes in the $a\rightarrow0$
limit. Since the mass shift in Fig.\ref{fig:mcrit} also disappears
with ${\rm O}(a^{2})$, such a shift does not invalidate this argument.
If the lines of constant residue in Fig.\ref{fig:scaling}a were quadratic,
a staggered action with $am_{{\rm st}}\propto a^{2}$ would have a
constant deviation from a chiral overlap action \emph{even in the
continuum limit}. A chiral continuum limit thus requires that the
staggered mass that corresponds to a constant matching residue vanishes
faster than quadratically.

Fig.\ref{fig:scaling}b shows the mass at $10\%$ matching residue
in a logarithmic scale. The dashed lines are proportional to $a^{4}$
and while they describe the two--flavor data quite well, for the rooted
action only the finest lattices are in agreement with $a^{4}$ scaling.
However, in both cases the scaling is faster than quadratic (dotted
line), thus making a chiral continuum limit possible. Note that this
is the only place where we see a qualitative difference between the
unrooted and rooted actions. The arguments from Sect.\ref{sec:The-continuum-limit}
offer a possible explanation. Since ${\rm det}(T)$ is non-local,
on the coarser lattices this might affect our finite volume results,
and only close to the continuum where the non--local terms become
irrelevant does this finite volume effect go away. Apparently these
non--local terms are more pronounced for the rooted action than for
the unrooted one.

Finally we illustrate the matching using a physical observable, the
topological susceptibility $\langle Q^{2}\rangle/z^{2}$, as this
observable is very sensitive to the sea quarks. We define the topology
through the zero--modes of the smeared overlap operator used in the
matching and evaluate it on gauge ensembles generated with two and
one flavor/taste staggered and overlap actions at various masses.%
\begin{figure}
\includegraphics[%
  scale=0.8]{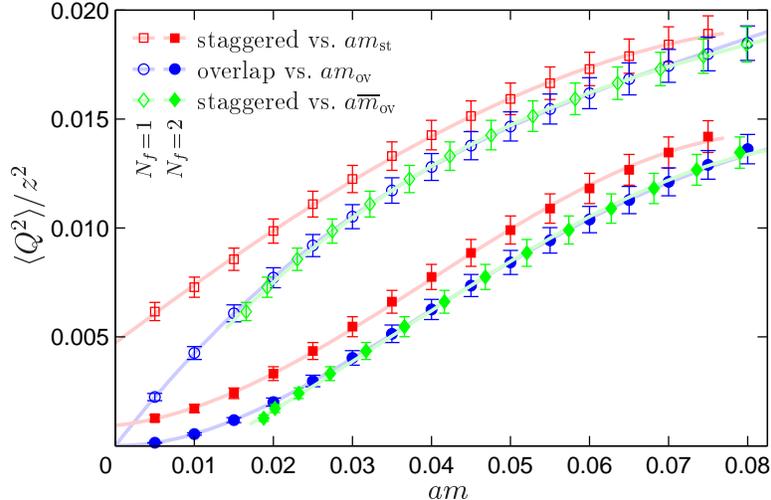}

\caption{The topological susceptibility on the $L/a=20$ ensemble. Open symbols
refer to the one and filled symbols to the two flavor/taste results.\label{fig:topo}
After shifting the staggered data (red squares) to the matched overlap
mass (green diamonds) almost perfect agreement with the overlap data
(blue circles) is achieved.}
\end{figure}
 Results from $L/a=20$ lattices are shown in Fig.\ref{fig:topo},
where the difference of the staggered and overlap ensemble at the
same bare fermion mass is very evident, especially at small masses
(red squares vs. blue circles). After shifting the staggered data
to the matching overlap mass (green diamonds), excellent agreement
is achieved. Again the one--flavor data shows better agreement since
in the reweighting to two flavors any remnant mismatch in the weight
becomes more important. The agreement on our finer lattices is equally
good and extends to smaller quark masses. Fig.\ref{fig:topo} is similar
to Fig.1 in Ref. \cite{Hasenfratz:2005ri} where matching between
the staggered and overlap data was attempted by a constant mass shift.
While a constant shift worked well for larger masses, it could not
reproduce the small quark region. The shift presented in Fig.\ref{fig:topo}
works everywhere but at the lightest mass.

\section{Conclusion}

The rooted staggered action is likely non--local in the physically
relevant range of small quark masses. However, this does not invalidate
the rooted action as long as the non--local terms are irrelevant and
scale away in the continuum limit. Here we demonstrated that this
is indeed the case in the 2--dimensional Schwinger model. We studied
how the staggered action differs from a chiral overlap action along
a line of constant physics as the continuum is approached. For both
the unrooted (as expected) and rooted staggered action we found that
the difference reduces to irrelevant local pure gauge terms. Nevertheless
care is required in taking the continuum limit of staggered fermions
such that the non QCD--like region is avoided.

It would be very interesting to repeat this calculation in 4 dimensions,
where the determinant ratios would have to be calculated numerically.
While difficult, it is not impossible to do that using stochastic
estimators and reduced determinants, especially in small volumes \cite{Hasenfratz:2005tt}.
Whether the matching is reliable on small volumes could be studied
within the Schwinger model first. 

\begin{acknowledgments}
A.H. benefited from discussions and presentations of the unpublished
works of Refs. \cite{Bernard:2006xx,Shamir:2006xx} during the ''Workshop
on the fourth root of the staggered fermion determinant'' at the INT
in Seattle and thanks the hospitality extended to her. We thank Y.
Shamir for discussions that helped clarify several of the points made
in this paper and also for a critical reading of the manuscript. We
thank Ulli Wolff and the Computational Physics Group at the Humboldt
University for the use of their computer resources. This research
was partially supported by the US Dept. of Energy.\bibliographystyle{apsrev}
\bibliography{lattice}

\end{acknowledgments}

\end{document}